\documentclass[showpacs,prc,twocolumn,floatfix,aps]{revtex4-1}
\usepackage{amssymb}
\usepackage{amsmath}
\usepackage{graphicx}
\usepackage{dcolumn}
\usepackage{amsfonts}
\usepackage{bm}
\newcommand{\bra}[1]{\left\langle #1\right|}
\newcommand{\ket}[1]{\left| #1\right\rangle}
\begin{document}

\title{Nonlocal extension of the dispersive-optical-model to describe data below the Fermi energy}
\author{W. H. Dickhoff$^{1}$, D. Van Neck$^{2}$, S. J. Waldecker$^{1}$, 
R. J. Charity$^{3}$, L. G. Sobotka$^{1,3}$}
\affiliation{Departments of Physics$^{1}$ and Chemistry$^{3}$, Washington University,
St.~Louis, Missouri 63130, USA
\\
Center for Molecular Modeling$^{2}$, Ghent University, Technologiepark 903, B-9052 Zwijnaarde, Belgium}
\date{\today }

\begin{abstract}
Present applications of the dispersive-optical-model analysis are restricted by the use of a local but energy-dependent version of the generalized Hartree-Fock potential.
This restriction is lifted by the introduction of a corresponding nonlocal potential without explicit energy dependence.
Such a strategy allows for a complete determination of the nucleon propagator below the Fermi energy with access to the expectation value of one-body operators (like the charge density), the one-body density matrix with associated natural orbits, and complete spectral functions for removal strength.
The present formulation of the dispersive optical model (DOM) therefore allows the use of elastic electron-scattering data in determining its parameters.
Application to ${}^{40}$Ca demonstrates that a fit to the charge radius leads to too much charge near the origin using the conventional assumptions of the functional form of the DOM.
A corresponding incomplete description of high-momentum components is identified, suggesting that the DOM formulation must be extended in the future to accommodate such correlations properly. 
Unlike the local version, the present nonlocal DOM limits the location of the deeply-bound hole states to energies that are consistent with (\textit{e,e}$^{\prime }$\textit{p}) and (\textit{p,2p}) data.
\end{abstract}

\maketitle

\section{INTRODUCTION}

\label{sec:intro}
An important link between the description of nuclear reactions and nuclear structure was proposed in Ref.~\cite{Mahaux86}, establishing a workable connection between the optical-model and the shell-model potential.
The proposed implementation is now known as the dispersive optical model (DOM) and has been extensively reviewed in Ref.~\cite{Mahaux91}.
Using empirical knowledge of standard optical potentials Mahaux and Sartor proposed to employ the experimentally well-constrained imaginary parts of the optical potential as the critical building blocks to determine, through a subtracted dispersion relation, the corresponding dynamic real part without requiring additional parameters.
Since the subtraction point is usually chosen to be the Fermi energy, the remaining static and real Hartree-Fock-like potential can be linked to empirical information from mean-field theories.
Conventional Woods-Saxon form factors have been employed including standard forms for volume and surface contributions.
By assuming a similar energy dependence of the imaginary part of the potential \textit{above} and \textit{below}
the Fermi energy, it is then possible to derive and successfully predict properties of bound nucleons, including removal energies, overlap functions, spectroscopic factors, and widths~\cite{Mahaux91}. 

The success in reproducing spectroscopic factors derived from the (\textit{e,e}$^{\prime }$\textit{p})  reaction~\cite{Lapikas93} relies partly on the inclusion of an energy asymmetry far from the Fermi energy for the imaginary volume contribution associated with the larger phase space of particle-like as compared to hole-like states.
This asymmetry can be inferred from nuclear-matter calculations of the imaginary part of the self-energy~\cite{Mahaux91}.
A number of studies have made use of the framework of the DOM~\cite{Mahaux87,Jeukenne88,Mahaux88,Mahaux89,Tornow90,Wang93,Molina02,Nagadi03,Chen04}.
Some steps towards a global version of the DOM have been recently reported in~\cite{Morillon04,Morillon07,Li08,BCDS10}.
A different perspective was recently developed in~\cite{Charity06,Charity07} where the DOM was exploited to extract the nucleon asymmetry dependence from fitting ${}^{40}$Ca and ${}^{48}$Ca data.
This allows both interpolation and extrapolation to other nuclei and the prediction of experimental data.
Such data-driven extrapolations may play an important role in predicting the properties of nuclei towards the drip lines.

Additional benefits may result when DOM ingredients are used in the description of transfer 
reactions using the adiabatic distorted-wave approximation which employs proton and neutron optical potentials for the description of the relevant deuteron scattering wave function~\cite{Johnson05,Zhanov07}.
The spectroscopic factors deduced from transfer reactions continue to exhibit a strong dependence on which optical potential is employed, as was recently shown for a number of Ar isotopes~\cite{Lee10}.
In addition to this uncertainty, there is a substantial discrepancy between the extracted single-particle (sp) properties from transfer and heavy-ion knockout reactions~\cite{Gade04,Gade08}.
Without a means to unambiguously extract such sp properties, the success of rare-isotope facilities will be severely hampered.
A well-constrained complete optical potential in the sense of the self-energy for the Dyson equation for nucleons will therefore be an important ingredient in obtaining unique and undisputed information~\cite{Dickhoff10} linking structure and reaction data in a unified manner.

The DOM can easily be applied to elastic-scattering data, since only phase shifts are required to describe differential cross sections and polarization data.
To be useful for transfer reactions, knowledge of the interior wave functions of protons and neutrons at positive energy is required.
The current implementation of the DOM employs a real component, a Hartree-Fock-like term, that is in principle nonlocal but is replaced by an equivalent local potential with an energy dependence mainly governed by the so-called \textit{k}-mass~\cite{Mahaux91}.
Since this energy dependence does not result from a dispersion integral, it leads to a distortion of the normalization that can be approximately fixed~\cite{Perey62,Fiedeldey66,Dickhoff08}.
Since specific assumptions about a gaussian form of the nonlocality are made in addition to the approximate nature of the local approximation, it is quite important to explore the actual form of nonlocal potentials that are based on microscopic calculations of the nucleon self-energy like in the Faddeev random phase approximation (FRPA)~\cite{Barb:1,Barb:2,Dickhoff04}.
Another reason to explore the inclusion of a nonlocal Hartree-Fock-like potential is to allow additional data to be included in the fitting procedure.
It is the goal of the present paper to explore the inclusion of a nonlocal Hartree-Fock-like potential in the DOM in order to describe a larger set of data in particular those pertaining to properties of nucleons below the Fermi energy.

The normalization distortion is particularly significant below the Fermi energy, where it leads to difficulties in determining the sp strength distribution~\cite{Mahaux91}. 
A nonlocal DOM potential can avoid these problems and provide a properly normalized solution of the Dyson equation.
The resulting propagator below the Fermi energy then provides access to spectral functions, the one-body density matrix, and all one-body expectation values in the ground state.
In turn, elastic electron-scattering data that yield the nuclear charge distribution can  be used to constrain the DOM potentials.
The sp strength distributions obtained from (\textit{p,2p})~\cite{Jacob66,Jacob73} and (\textit{e,e}$^{\prime }$\textit{p})~\cite{Lapikas93} reactions provide further constraints.
A nonlocal DOM potential also avoids the pitfall of the linear energy dependence of the equivalent local potential which provides too much binding for the most deeply bound levels. 

Experimental evidence for the presence of high-momentum components in the nuclear ground state has been provided by the (\textit{e,e}$^{\prime }$\textit{p}) reaction~\cite{Rohe04} in reasonable agreement with microscopic calculations for light nuclei~\cite{Muther94,Muther95}.
While the number of high-momentum protons in light nuclei represents a modest 10\% of the total, their presence does confirm the basic tenet of most realistic nucleon-nucleon interactions that contain a sizable repulsion at short relative distances.
It is unclear whether current DOM implementations generate any sizable presence of high-momentum nucleons.
It is therefore useful to explore the spectral distribution in momentum space of the DOM propagator to assess its high-momentum content, and, if found lacking, provide an incentive to construct DOM self-energies that represent the experimental findings~\cite{Rohe04}.
While spectroscopic factors obtained from the analysis of the (\textit{e,e}$^{\prime }$\textit{p}) reactions are already employed in current DOM fits~\cite{Charity06,Charity07}, a nonlocal treatment of the DOM potential opens the possibility to use these cross-section data directly in the fitting procedure, since both the overlap function of the removed proton as well as the scattering wave function of the outgoing proton can be described by the corresponding DOM wave functions.
Data from the (\textit{e,e}$^{\prime }$\textit{p}) reaction therefore can provide additional constraints and provide further confirmation of the interpretation of these data as well as reducing the uncertainty in the quoted absolute spectroscopic factors~\cite{Dickhoff10}.

The purpose of this work is to clarify the inclusion of a nonlocal Hartree-Fock-like potential and generate the resulting solution of the Dyson equation below the Fermi energy, while keeping the earlier DOM results obtained with the local equivalent potential intact or improving upon them.
We will continue to make the simplest possible assumptions about the form of the nonlocality and thereby not introduce any additional parameters.
In Sec.~\ref{Sec:formalism} we will clarify the required strategy to solve the Dyson equation in coordinate space while preserving the ingredients obtained from earlier fits to ${}^{40}$Ca data~\cite{Charity06,Charity07}.
The results illustrating the complete solution of the Dyson equation below the Fermi energy are illustrated and discussed in Sec.~\ref{Sec:results}.
A summary and conclusions are presented in Sec.~\ref{Sec:conclusions}.

\section{Dyson equation and DOM self-energy with nonlocal potentials}

\label{Sec:formalism} 
\subsection{Green's function ingredients}
\label{sec:GF}
We start with a brief summary of relevant results from the Green's function formulation of the many-body problem~\cite{Dickhoff08}.
The nucleon propagator with respect to the $A$-body ground state is given by
\begin{eqnarray}
G _{\ell j}(r ,r' ; E) 
&=  \sum_m \frac{\bra{\Psi^A_0} a_{r\ell j}
\ket{\Psi^{A+1}_m} \bra{\Psi^{A+1}_m} a^\dagger_{r' \ell j} \ket{\Psi^A_0}
}{ E - (E^{A+1}_m - E^A_0 ) +i\eta }  & \nonumber \\
& +  \sum_n \frac{\bra{\Psi^A_0} a^\dagger_{r' \ell j} \ket{\Psi^{A-1}_n}
\bra{\Psi^{A-1}_n} a_{r \ell j} \ket{\Psi^A_0} }{
E - (E^A_0 - E^{A-1}_n) -i\eta} , &
\label{eq:prop}
\end{eqnarray}
where complete sets of states in the $A\pm1$ systems are inserted and the sp basis with good radial position, orbital angular momentum (parity) and total angular momentum is chosen while suppressing the projection of the total angular momentum and the isospin quantum numbers.
The continuum solutions in the $A\pm1$ systems are also implied in the completeness relations.
The numerators of the particle and hole components of the propagator represent the products of overlap functions associated with adding or removing a nucleon from the $A$-body ground state.
The standard development of Green's function theory relates the nucleon propagator to the self-energy yielding the Dyson equation in the following form
\begin{eqnarray}
\label{eq:dyson}
&G_{\ell j}(r,r';E) = G^{(0)}_{\ell j}(r,r';E) & \\
&+ \int \!\! d\tilde{r}\ \tilde{r}^2 \!\! \int \!\! d\tilde{r}'\ \tilde{r}'^2 G^{(0)}_{\ell j}(r,\tilde{r};E)
\Sigma_{\ell j}(\tilde{r},\tilde{r}';E) G_{\ell j}(\tilde{r}',r';E) . &
\nonumber
\end{eqnarray}
For the present discussion the noninteracting propagator involves only kinetic energy contributions.
The nucleon self-energy contains all linked diagrammatic contributions that are irreducible with respect to propagation represented by $G^{(0)}$.
All contributions to the propagator are then generated by the Dyson equation itself.
The solution of the Dyson equation generates all discrete poles corresponding to bound $A\pm1$ states explicitly given by Eq.~(\ref{eq:prop}) that can be reached by adding or removing a particle with quantum numbers $r \ell j$.
The hole spectral function is obtained from
\begin{equation}
S_{\ell j}(r;E) = \frac{1}{\pi}  \textrm{Im}\ G_{\ell j}(r,r;E)  
\label{eq:holes}
\end{equation}
for energies in the continuum.
The total spectral strength at $E$ for a given $\ell j$ combination, 
\begin{equation}
S_{\ell j}(E) = \int_{0}^\infty dr\ r^2\ S_{\ell j}(r;E) ,
\label{eq:specs}
\end{equation}
yields the spectroscopic strength per unit of energy.
For discrete energies as well as all continuum ones, overlap functions for the addition or removal of a particle are generated as well.
The connection between the nucleon propagator and elastic-scattering data can therefore be made explicit by identifying the nucleon elastic-scattering $\mathcal{T}$-matrix with the reducible self-energy obtained by iterating the irreducible one to all orders with $G^{(0)}$~\cite{Bell59,Villars67, BlaizotR86,Dickhoff08}.

For discrete states in the $A-1$ system one can show that the overlap function obeys a Schr{\"o}dinger-like equation~\cite{Dickhoff08}.
Introducing the notation
\begin{equation}
\psi^n_{\ell j}(r) = \bra{\Psi^{A-1}_n}a_{r \ell j} \ket{\Psi^A_0} ,
\label{eq:overlap}
\end{equation}
for the overlap function for the removal of a nucleon at $r$ with discrete quantum numbers $\ell$ and $j$, one finds
\begin{eqnarray}
\left[ \frac{ p_r^2}{2m} +
 \frac{\hbar^2 \ell (\ell +1)}{2mr^2}\right]  & \psi^{n}_{\ell j}(r) & \nonumber  \\
+   \int \!\! dr'\ r'^2 
\Sigma_{\ell j}(r,r';\varepsilon^-_n)  &\psi^{n}_{\ell j}(r') & = 
\varepsilon^-_n \psi^{n}_{\ell j}(r) ,
\label{eq:DSeq}
\end{eqnarray}
where
\begin{equation}
\varepsilon^-_n=E^A_0 -E^{A-1}_n 
\label{eq:eig}
\end{equation}
and in coordinate space the radial momentum operator is given by $p_r = -i\hbar(\frac{\partial}{\partial r} + \frac{1}{r})$.
Discrete solutions to Eq.~(\ref{eq:DSeq}) exist in the domain where the self-energy has no imaginary part and these are normalized by utilizing the inhomogeneous term in the Dyson equation.
For an eigenstate of the Schr{\"o}dinger-like equation [Eq.~(\ref{eq:DSeq})], the so-called quasihole state labeled by $\alpha_{qh}$, the corresponding normalization or spectroscopic factor is given  by~\cite{Dickhoff08}
\begin{equation}
S^n_{\ell j} = \bigg( {1 - 
\frac{\partial \Sigma_{\ell j}(\alpha_{qh},
\alpha_{qh}; E)}{\partial E} \bigg|_{\varepsilon^-_n}} 
\bigg)^{-1} ,
\label{eq:sfac}
\end{equation}
which is the discrete equivalent of Eq.~(\ref{eq:specs}).
Discrete solutions in the domain where the self-energy has no imaginary part can therefore be obtained by expressing Eq.~(\ref{eq:DSeq}) on a grid in coordinate space and performing the corresponding matrix diagonalization. 
Likewise, the solution of the Dyson equation [Eq.~(\ref{eq:dyson})] for continuum energies in the domain below the Fermi energy, can be formulated as a complex matrix inversion in coordinate space.
This is advantageous in the case of a nonlocal self-energy representative of all microscopic approximations that include at least the Hartree-Fock approximation.
Below the Fermi energy for the removal of a particle
\begin{equation}
\varepsilon^-_F = E^A_0 -E^{A-1}_0 ,
\label{eq:efmin}
\end{equation}
the corresponding discretization is limited by the size of the nucleus as can be inferred from the removal amplitude given in Eq.~(\ref{eq:overlap}), which demonstrates that only coordinates inside the nucleus need to be considered.
Such a finite interval therefore presents no numerical difficulty.

\subsection{Link with the DOM}
\label{sec:GFDOM}
While microscopic calculations of the nucleon self-energy have made substantial progress in recent years~\cite{Barb:1,Barb:2,Dickhoff04,Barb09}, accurate representations of elastic-scattering data in particular have not yet gone beyond phenomenological representations of the self-energy in terms of standard or dispersive optical potentials.
A clear link with the microscopic self-energy is provided by the DOM strategy~\cite{Mahaux86,Mahaux91}.
It employs the dispersion relation between the real and imaginary part of the microscopic self-energy given by
\begin{eqnarray} 
\mbox{Re} \Sigma_{\ell j}(r,r';E)\!& =& \! \Sigma^s_{\ell j} (r,r')\! - \! {\cal P} \!\!
\int_{\varepsilon_T^+}^{\infty} \!\! \frac{dE'}{\pi} \frac{\mbox{Im} \Sigma_{\ell j}(r,r';E')}{E-E'}  \nonumber \\
&+&{\cal P} \!\!
\int_{-\infty}^{\varepsilon_T^-} \!\! \frac{dE'}{\pi} \frac{\mbox{Im} \Sigma_{\ell j}(r,r';E')}{E-E'} ,
\label{eq:disprel}
\end{eqnarray}
where $\mathcal{P}$ represents the principal value.
The static contribution arises from the correlated Hartree-Fock term and the dynamic parts start and end at corresponding thresholds in the $A\pm1$ systems that have a larger separation than the corresponding difference between the Fermi energies for addition ($\varepsilon_F^+ = E^{A+1}_0-E^A_0$) and removal ($\varepsilon_F^-=E^A_0-E^{A-1}_0$) of a particle.
The latter feature is particular to a finite system and generates possibly several discrete quasiparticle and hole-like solutions of the Dyson equation in Eq.~(\ref{eq:DSeq}) in the domain where the imaginary part of the self-energy vanishes.

The standard definition of the self-energy requires that its imaginary part is negative, at least on the diagonal, in the domain that represents the coupling to excitations in the $A+1$ system, while it is positive for the coupling to $A-1$ excitations.
This translates into an absorptive potential for elastic scattering at positive energy, where the imaginary part is responsible for the loss of flux in the elastic channel.
It is convenient to introduce the average Fermi energy
\begin{equation}
\varepsilon_F = \frac{1}{2} \left[
\varepsilon_F^+  - \varepsilon_F^- \right] .
\label{eq:FE}
\end{equation}
Subtracting Eq.~(\ref{eq:disprel}) calculated at this energy, from Eq.~(\ref{eq:disprel}) generates the so-called subtracted dispersion relation 
\begin{eqnarray} 
\!\!\!\!\! \mbox{Re} \Sigma_{\ell j}(r,r';E)\! = \!  \mbox{Re} \Sigma_{\ell j} (r,r';\varepsilon_F) \hspace{2.0cm}  \nonumber \\
- \! {\cal P} \!\!
\int_{\varepsilon_T^+}^{\infty} \!\! \frac{dE'}{\pi} \mbox{Im} \Sigma_{\ell j}(r,r';E') \left[ \frac{1}{E-E'}  - \frac{1}{\varepsilon_F -E'} \right]  \nonumber  \\
+{\cal P} \!\!
\int_{-\infty}^{\varepsilon_T^-} \!\! \frac{dE'}{\pi} \mbox{Im} \Sigma_{\ell j}(r,r';E') \left[ \frac{1}{E-E'}
-\frac{1}{\varepsilon_F -E'} \right]  .
\label{eq:sdisprel}
\end{eqnarray}
The beauty of this representation was recognized by Mahaux and Sartor~\cite{Mahaux86,Mahaux91} since it allows for a link with empirical information both at the level of the real part of the nonlocal self-energy at the Fermi energy (probed by a multitude of Hartree-Fock calculations) and also through empirical knowledge of the imaginary part of the optical potential (constrained by experimental data) that consequently yields a dynamic contribution to the real part by means of Eq.~(\ref{eq:sdisprel}).
This procedure requires further assumptions since detailed knowledge of the imaginary part of the self-energy below the Fermi energy has only become available with electron-induced proton knockout reactions~\cite{Lapikas93}.
Since the empirical knowledge has relied on local representations of the imaginary part of the optical potential, it is natural to make a similar assumption for the DOM version. 
In addition, a separation in terms of surface (low energy) and volume (higher energy) absorption has been incorporated in accordance with standard practice.

Most implementations of the DOM simply assume that near the Fermi energy there is a similar behavior above and below the Fermi energy for the imaginary part of the surface contribution to the self-energy although this assumption requires more stringent tests.
Farther away from the Fermi energy the phase space asymmetry characterized by the density of two-particle--one-hole (2p1h) and two-hole--one-particle (2h1p) states leads to an assumed energy asymmetry for the volume contribution that is also consistent with expectations from nuclear-matter results.
Standard forms and fitted parameters for a recent implementation of the DOM for Ca isotopes can be found in Ref.~\cite{Charity07}.

A clear success of the DOM approach is the accurate prediction of bound sp properties including radii, spectroscopic factors, and the compression of sp levels near the Fermi energy that originates from the dispersive contribution in Eq.~(\ref{eq:sdisprel})~\cite{Mahaux91} and cannot be obtained with standard Hartree-Fock approaches.
The latter feature provides an explanation for the observed nuclear level density parameter $a$ which is large compared to sp estimates~\cite{Bohr69}.

The remaining ingredient in the DOM procedure is the nonlocal self-energy at the Fermi energy represented by $\mbox{Re} \Sigma_{\ell j} (r,r';\varepsilon_F)$.
It is more convenient to discuss the spin-independent part of this term in a basis with vectors in coordinate space using the notation $\Sigma_{HF}(\bm{r},\bm{r}')$ and employing the HF label that was introduced by Mahaux and Sartor~\cite{Mahaux91} even though this term is not a true Hartree-Fock contribution as the derivation of Eq.~(\ref{eq:sdisprel}) clarifies. 
The usual treatment of $\Sigma_{HF}(\bm{r},\bm{r}')$ is to assume that it can be replaced by a local but energy-dependent potential~\cite{Perey62,Fiedeldey66,Mahaux91,Dickhoff08}.
The corresponding form then can be written as
\begin{equation}
\Sigma_{HF}(\bm{r},\bm{r}') \Rightarrow \mathcal{V}_{HF}(r,E) \delta(\bm{r}-\bm{r}') ,
\label{eq:loceq}
\end{equation}
where 
\begin{equation}
\mathcal{V}_{HF}(r,E) = V_{HF} (E)f(r,r_{HF},a_{HF})
\label{eq:HFmstar}
\end{equation}
containing the Woods-Saxon form factor
\begin{equation}
f(r,r_{i},a_{i})=\frac{1}{1+\exp({\frac{r-r_{i}A^{1/3}}{a_{i}}})}  .
\label{eq:WSf}
\end{equation}%
The factorized linear energy dependence can be parametrized by the corresponding effective mass below the Fermi energy and can be represented by
\begin{equation}
V_{HF}(E) =  V_{HF}(\varepsilon_F)+\left[1-\frac{m^*_{HF}}{m}\right] \left(E-\varepsilon_F\right) ,
\label{eq:HFEdep}
\end{equation}
which can be combined with the Woods-Saxon form factor to generate $m^*_{HF}(r)$.
This version is inspired by the Skyrme implementation of the HF potential~\cite{Mahaux91}.
More generally, one may identify this effective mass with an energy-dependent version of the effective mass $\widetilde{m}^*(r;E)$ that governs the nonlocality of the self-energy and is sometimes referred to as the $k$-mass.
It was shown in Ref.~\cite{NY81} that this effective mass is critical to reconcile the phenomenological (local) imaginary part of the optical potential with the microscopic one~\cite{Dickhoff08} and to explain the observed nucleon mean free path.
For finite nuclei, this implies that the DOM version of its local imaginary part $\mathcal{W}$ is related to the self-energy by
\begin{equation}
\mathcal{W}(r;E) = \frac{\widetilde{m}^*(r;E)}{m} \mbox{Im} \Sigma(r;E) .
\label{eq:effmass}
\end{equation}
This suggests that the use of a nonlocal HF self-energy in the DOM framework has to be accompanied by enhancing the imaginary part with a corresponding factor $m/\widetilde{m}^*(r;E)$.
Results discussed later indeed corroborate the necessity of including this factor \textit{e.g.} to obtain identical spectroscopic factors as those in Ref.~\cite{Charity07}.
It is therefore possible to employ the same parameters as in the fit of~\cite{Charity07} and only replace the energy-dependent local equivalent HF potential by a suitable energy-independent nonlocal one.
We have chosen the standard form introduced in Ref.~\cite{Perey62} to represent
\begin{equation}
\Sigma_{HF}(\bm{r},\bm{r}') = V_{NL} f({\scriptstyle{\frac{1}{2}}}|\bm{r}+\bm{r}'|,r_{NL},a_{NL}) H(\left|\bm{r}-\bm{r}'\right|) ,
\label{eq:NL}
\end{equation}
where the degree of nonlocality is expressed by a gaussian governed by the parameter $\beta$
\begin{equation}
H(|\bm{r}-\bm{r}'|) = \frac{1}{\pi^{\frac{3}{2}} \beta^3} 
 \exp{\left[ -\left(\frac{\bm{r}-\bm{r}'}{\beta}\right)^2 \right]} .
\label{eq:beta}
\end{equation}
This nonlocal form requires four parameters ($V_{NL},r_{NL},a_{NL}$, and $\beta$), which is the same number required to represent $\mathcal{V}_{HF}(r,E)$ in Eq.~(\ref{eq:HFmstar}).
We reiterate that this nonlocal representation is essential in obtaining properly normalized spectral functions and spectroscopic factors.
In the following we will discuss results for ${}^{40}$Ca with this nonlocal version of the DOM with emphasis on energies below the Fermi energy.
We note that all DOM parameters that were obtained in Ref.~\cite{Charity07} have been kept while only the local HF potential with its spurious energy dependence has been replaced by Eq.~(\ref{eq:NL}) and the application of the dispersion relation Eq.~(\ref{eq:sdisprel}) has been modified according to Eq.~(\ref{eq:effmass}).

\section{Results}

As discussed in the previous section, only four parameters are needed to introduce the nonlocal HF contribution to the DOM.
As in the usual DOM fit, the location of the main fragments of the $0d_{3/2}$ and $1s_{1/2}$ valence hole levels was used to constrain the parameters of the nonlocal HF potential.
Since the complete one-body density matrix can be obtained with a nonlocal HF potential, it was also possible to constrain the parameters by the mean square radius of the charge distribution that is well known experimentally~\cite{deVries1987}.
\label{Sec:results} 
\begin{table}[bp]
\caption{Parameters for the local energy-dependent Woods-Saxon potential and the nonlocal version with gaussian nonlocality for ${}^{40}$Ca.}
\label{Tbl:parms}%
\begin{ruledtabular}
\begin{tabular}{ccc}
 & local &nonlocal\\ 
\hline
Depth [MeV] & -56.5 & -92.3 \\
Radius [fm] & 1.19 & 1.05 \\
Diffuseness [fm]& 0.70 & 0.70 \\
$\widetilde{m}^*_{HF}/m$ & 0.57 & -\\
Nonlocality [fm] & -& 0.91 \\
\end{tabular}
\end{ruledtabular}
\end{table}
An additional problem that can be cured by the nonlocal version of the HF potential is associated with the linear energy dependence of the local version as shown in Eq.~(\ref{eq:HFEdep}).
Typical DOM fits generate rather deeply bound $0s_{1/2}$ states, often well below the peaks seen in (\textit{e,e}$^{\prime }$\textit{p}) and (\textit{p,2p}) experiments.
With a nonlocal potential it is possible to use the peak of the deeply bound $0s_{1/2}$ state as an additional constraint and avoid the problem.
The resulting parameters are collected in Table~\ref{Tbl:parms} including those for the local potential.
All other parameters and the detailed shapes chosen for the imaginary parts of the DOM potential for ${}^{40}$Ca can be found in Ref.~\cite{Charity07}.
When adjusting the parameters of the nonlocal potential it was found that it was possible to incorporate the constraint of the mean square radius of the charge distribution while generating quasihole fragments at energies that are at least as good as the original fit.
It may be useful in the future to explore other forms for nonlocal potentials, especially when microscopic self-energies obtained with the FRPA method~\cite{Barb:1,Barb:2,Dickhoff04,Barb09} are analyzed.
It is possible to keep the same value for the diffuseness parameter as for the local potential.
Depth and radius parameters are however clearly very different and must combine with the nonlocality parameter $\beta$ to generate similar results for the energies of the valence hole states. 
It was found that the charge radius of the nucleus provides a significant constraint and should therefore be used in future applications of the DOM method.

\begin{figure}[tbp]
\includegraphics*[width=.37\textwidth,angle=-90]{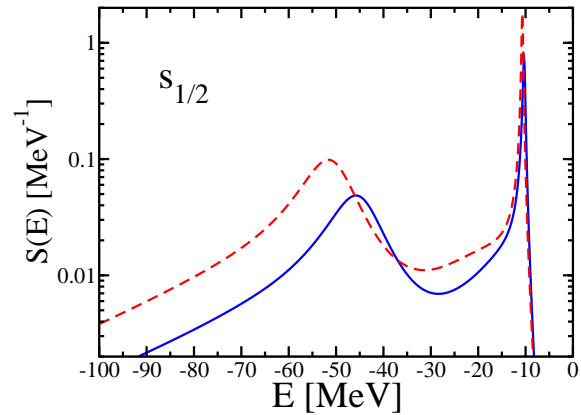}
\caption{(Color online) Comparison of proton $s_{1/2}$ spectral strength with the nonlocal (solid) and local potential with the spurious energy dependence (dashed). The nonlocal result does not yet contain the correction given in Eq.~(\protect\ref{eq:effmass}).
Note that the dashed curve even exceeds the number of mean-field $s_{1/2}$ particles by more than 50\% illustrating the incorrect normalization when the local energy-dependent potential is used in the Dyson equation without proper corrections.}
\label{fig:s12bob}
\end{figure}

After projecting the nonlocal potential on to states with good orbital angular momentum, it is possible to perform the complex matrix inversion in coordinate space to solve Eq.~(\ref{eq:dyson}).
We note that the imaginary part of the DOM potential of Refs.~\cite{Charity06,Charity07} ends at the Fermi energy [see Eq.~(\ref{eq:FE})] so this procedure generates sharply peaked features for valence hole states just below the Fermi energy.
To illustrate the influence of the spurious energy dependence included for a local potential,  we display in Fig.~\ref{fig:s12bob} the proton $s_{1/2}$ spectral strength [see Eq.~(\ref{eq:specs})] with a nonlocal potential (solid) and the local version (dashed) from the original fit. 
The nonlocal potential yields the same valence quasihole energy but the correction implied by Eq.~(\ref{eq:effmass}) was not yet applied.
Spectral functions obtained by solving the Dyson equation with the spurious energy dependence associated with the local HF potential typically overestimate the mean-field occupation by more than 50\% leading to more than 30 protons for ${}^{40}$Ca.

\begin{figure*}[tbp]
\includegraphics*[width=0.68\textwidth,angle=-90]{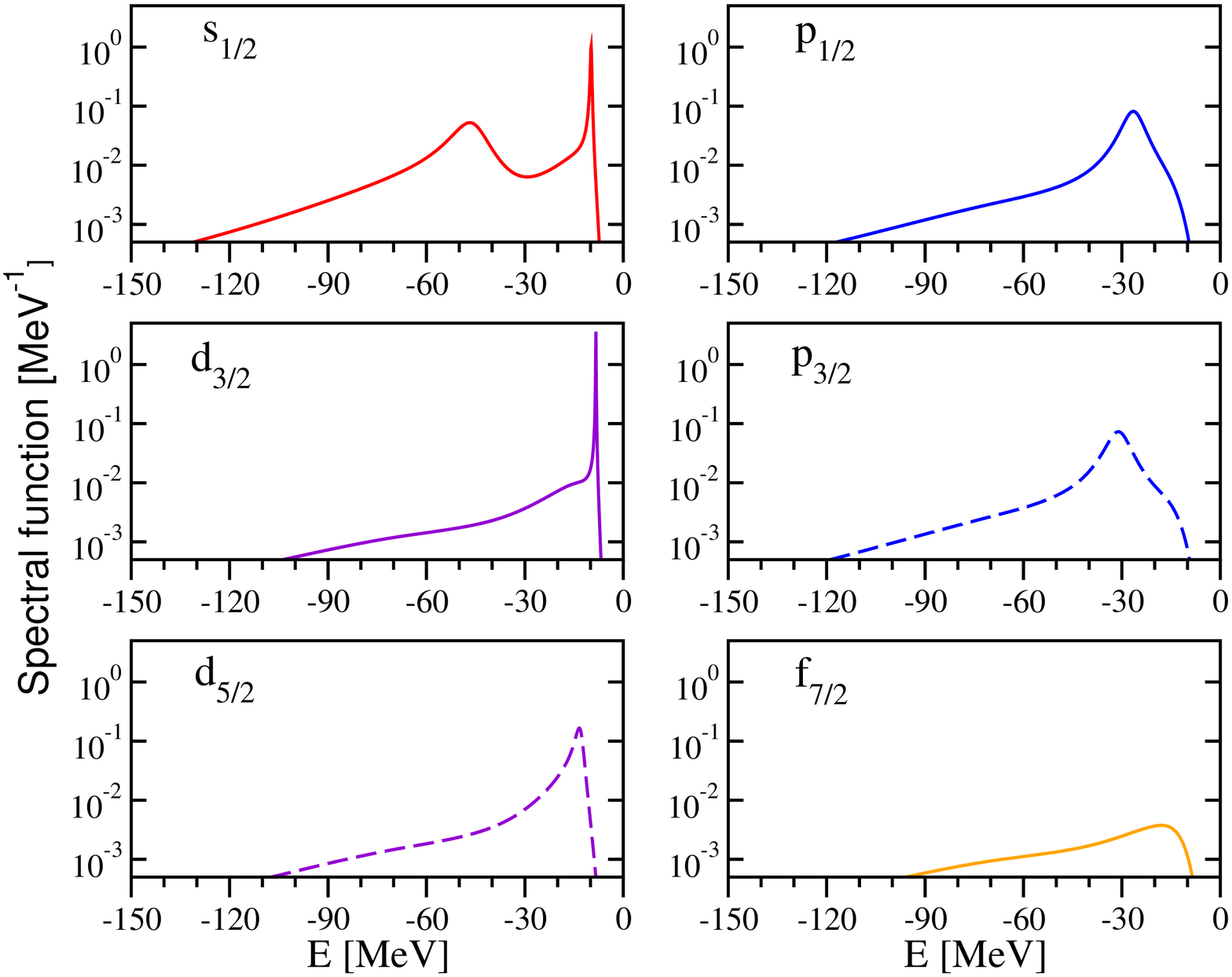}
\caption{(Color online) Spectral functions for all mostly occupied $\ell j$ combinations in ${}^{40}$Ca} together with the $f_{7/2}$ result. Orbits with $j = \ell +\frac{1}{2}$ are dashed to distinguish them from the ones with $j= \ell - \frac{1}{2}$. These results exhibit similar peak locations and widths as observed in (\textit{p,2p})~\protect\cite{Jacob66,Jacob73} and (\textit{e,e}$^{\prime }$\textit{p}) experiments~\protect\cite{FM84}.
\label{fig:allsf}
\end{figure*}
While general features of the solid curve in Fig.~\ref{fig:s12bob} for $s_{1/2}$ strength distribution appear in order, like the wide peak for the lowest orbit and a sharp well-localized peak for the one near the Fermi energy, it  should be noted that the strength in the peak near the Fermi energy contains a spectroscopic strength of 0.78 not in agreement with the DOM results of Refs.~\cite{Charity06,Charity07}.
It is only when the remainder of the DOM potential is multiplied by the effective mass correction of Eq.~(\ref{eq:effmass}) that we obtain a spectroscopic factor of 0.65 for the $1s_{1/2}$ state in better agreement with Ref.~\cite{Charity07}.
The correct strength distributions for the most relevant proton $\ell j$ combinations including $f_{7/2}$ are displayed in Fig.~\ref{fig:allsf}.
All peaks correspond to the orbits that are expected to be fully occupied in the mean field.
It is important to realize however, that at each energy the total strength according to Eq.~(\ref{eq:specs}) is obtained for a given $\ell j$ combination.
For a given (reasonable) mean field potential, several of the corresponding orbits may exhibit finite amounts of strength at one energy, including those that are not occupied in that mean field.
The converse is illustrated by the $f_{7/2}$ strength distribution, since this orbit is completely empty in a mean-field picture. 
The strength exhibited appears on account of the presence of the imaginary part of the self-energy below the Fermi energy which allows some finite amount of $f_{7/2}$ strength to appear there.
In addition to this feature, mostly occupied $\ell j$ combinations exhibit a broadening of the strength with a width that represents the local mixing with more complicated states like 2h1p, \textit{etc.} 
We note that it is straightforward to place the first peak of the $s_{1/2}$ distribution in accordance with experimental observations from (\textit{p,2p})~\cite{Jacob66,Jacob73} and (\textit{e,e}$^{\prime }$\textit{p}) experiments~\cite{FM84}.
This is more difficult when the local HF potential is employed due to its linear energy dependence which lowers the well with decreasing energy.
When integrating the total strength shown in Fig.~\ref{fig:allsf} for all orbits except the $f_{7/2}$ and multiplying with the corresponding degeneracy factor of $2j+1$, the summed strength is 19.48.
While this may appear reasonable, it should be kept in mind that the assumed state independence of the DOM potential (apart from spin-orbit) and the $\ell$-dependence of the nonlocal HF potential (on account of its angular dependence implied by Eq.~(\ref{eq:NL})), implies that some strength will also be generated for higher $\ell$-values leading to an overestimate of the total proton number. 
Indeed, when the cut-off is placed at $\ell =3$, \textit{i.e.} the $f_{7/2}$ and $f_{5/2}$ contributions are also included, the total proton number becomes 21.43. 
This suggests that in future DOM work the total proton (neutron) number should be used as a further constraint on the potentials.
The possibility of including some state dependence may also be explored, in particular by relying on microscopic input from FRPA calculations~\cite{Barb:1,Barb:2,Dickhoff04,Barb09}.

Before discussing new results not available with the standard DOM implementation, we first compare several quasiparticle properties in the two approaches.
In Table~\ref{Tbl:qpprop} we compare quasihole energies obtained with the local and nonlocal DOM with experimental data (for deeply-bound orbits we use Ref.~\cite{Jacob73}).
The numbers quoted in the following tables for quasiparticle properties using the local version of the DOM differ slightly from the ones generated in Ref.~\cite{Charity07}, since a small error in the calculation of the dispersive volume contribution has been corrected.
\begin{table}[tbp]
\caption{Quasihole energies for proton orbits in ${}^{40}$Ca for the local and nonlocal DOM implementation compared with experiment.}
\label{Tbl:qpprop}%
\begin{ruledtabular}
\begin{tabular}{ccccc}
 & & Energy [MeV] & \\ 
orbit  &local&nonlocal & peak & experiment\\
\hline
$0s_{1/2}$ & -57.3 & -47.4 & -46.7 & $\sim$-47\\
$0p_{3/2}$ & -35.1 & -31.4 & -31.1 & $\sim$-30\\
$0p_{1/2}$& -30.3 & -26.7 & -26.4 & $\sim$-30\\
$0d_{5/2}$ & -13.5 & -13.8 & -13.5 & -13.5\\
$1s_{1/2}$ & -9.5 & -9.8 & -9.8 & -10.8 \\
$0d_{3/2}$ & -8.3 & -8.3 & -8.3 & -8.3 
\end{tabular}
\end{ruledtabular}
\end{table}
The column labeled ``local'' reports the solutions of the eigenvalue equation for the local DOM potential without the imaginary part.
This includes a self-consistency procedure since the potential is energy dependent, \textit{i.e.} the chosen input energy has to coincide with the obtained eigenvalue.
Such a calculation for the nonlocal DOM is reported in the column labeled nonlocal.
The imaginary part is included in the column labeled ``peak'' which identifies the location of the peak of the spectral function for each orbit.
As expected, there is little difference between the latter two approaches, especially close to the Fermi energy.
The largest difference between the local and nonlocal approach occurs for the lowest $s_{1/2}$ orbit.
As discussed earlier, the nonlocal potential is better able to constrain the peak of the spectral $s_{1/2}$ strength to the correct value.
Overall agreement for the peak location appears quite satisfactory for the nonlocal DOM although it may be necessary to consider some state dependence if a better fit for $1s_{1/2}$ quasihole energy is deemed appropriate.
The results for neutrons are naturally not very different apart from the Coulomb shift to those for protons on account of isospin symmetry and are therefore not reported.
We only note that the lowest neutron $s_{1/2}$ peak occurs at -56 MeV in the nonlocal DOM while in the local counterpart it is found at -67 MeV, confirming the discussion for the corresponding proton level that the local DOM tends to bind this orbit too deeply.

In Table~\ref{Tbl:spfac} the spectroscopic factors are listed for the same orbits as in Table~\ref{Tbl:qpprop}.
\begin{table}[btp]
\caption{Spectroscopic factors for proton orbits in ${}^{40}$Ca for the local and nonlocal DOM implementation.}
\label{Tbl:spfac}%
\begin{ruledtabular}
\begin{tabular}{ccc}
orbit  &local&nonlocal \\
\hline
$0s_{1/2}$ & 1.11 & 0.98 \\
$0p_{3/2}$ & 0.94 & 0.93 \\
$0p_{1/2}$& 0.95 & 0.94 \\
$0d_{5/2}$ & 0.83 & 0.86\\
$1s_{1/2}$ & 0.67 & 0.65 \\
$0d_{3/2}$ & 0.65 & 0.64 
\end{tabular}
\end{ruledtabular}
\end{table}
These results are obtained for the local DOM by using the approximate expression for the spectroscopic factor reviewed in Ref.~\cite{Mahaux91} and given explicitly in Eq.~(12) of Ref.~\cite{Charity07}.
This expression does not guarantee that the resulting spectroscopic factor is less than 1 (as it should be), which is illustrated by the outcome for the $0s_{1/2}$ orbit.
For the nonlocal DOM, Eq.~(\ref{eq:sfac}) has been used where the derivative is taken at the eigenvalue obtained from Eq.~(\ref{eq:DSeq}) with neglect of the imaginary part of the potential.
This procedure is also not appropriate in the domain where the imaginary part becomes substantial and is already suspect for the $d_{5/2}$ orbit.
When the imaginary part is neglected, it is possible that the total real dispersive correction has a positive derivative at the energy corresponding to the self-consistent eigenvalue even in the nonlocal case, leading to an unphysical spectroscopic factor.  
Already the strength content of the peak for the $d_{5/2}$ orbit in Fig.~\ref{fig:allsf} is more in line with the spectroscopic factors quoted for the $1s_{1/2}$ and $0d_{3/2}$ orbits and therefore substantially smaller than the 0.86 listed in Table~\ref{Tbl:spfac}.
Only for the latter two orbits is the neglect of the imaginary part of the potential unimportant, since the content of the sharp peaks in Fig.~\ref{fig:allsf} coincides with the spectroscopic factors given in Table~\ref{Tbl:spfac}. 
We also note that there is reasonable agreement with the local and nonlocal DOM results for these levels.
We therefore conclude that only for these orbits the use of spectroscopic factors is sensible and unambiguous. 
This is in complete accord with the notion that the Landau quasiparticle (hole) concept is only valid in the immediate vicinity of the Fermi energy~\cite{Lan57a,Lan57b,Lan59} which is discussed for nuclei \textit{e.g.} in Refs.~\cite{Brand91,Dickhoff98}.
As discussed in Ref.~\cite{Charity07}, these quasihole spectroscopic factors are consistent with the analysis of the (\textit{e,e}$^{\prime }$\textit{p}) reaction on this nucleus for the $d_{3/2}$ orbit~\cite{Kramer89,Kramer01}.
These references quote a result for the $1s_{1/2}$ orbit that is too small and has been superseded by more recent experiments~\cite{Lapikas05}.
For more deeply-bound orbits, including the $d_{5/2}$, it is much more appropriate to consider the complete spectral functions as shown in Fig.~\ref{fig:allsf}.
We note that the very sharp peaks in the immediate vicinity of the Fermi energy can also be replaced by delta-functions with the strength given by the spectroscopic factor.

By integrating the imaginary part of the propagator given in Eq.~(\ref{eq:prop}) for each $\ell j$ combination up to the Fermi energy it is possible to obtain the one-body density matrix element
\begin{eqnarray}
n_{\ell j}(r',r) & = &  \frac{1}{\pi} \int_{-\infty}^{\varepsilon_F} dE\ \mbox{Im} G_{\ell j}(r,r';E) \nonumber \\
& = &
\bra{\Psi^A_0}a^\dagger_{r' \ell j} a_{r\ell j} \ket{\Psi^A_0} .
\label{eq:dmat}
\end{eqnarray}
For protons, the point charge distribution is thus obtained from the diagonal matrix elements of the one-body density matrix
\begin{equation}
\rho_p(r) = \frac{e}{4\pi} \sum_{\ell j} (2j+1) n_{\ell j}(r,r) .
\label{eq:chd}
\end{equation}
For a comparison with the experimental charge density of ${}^{40}$Ca it is necessary to fold this distribution with the proton charge density.
We used the procedure outlined in Ref.~\cite{Brown79} which employs 3 gaussians for the proton.
The mean square radius of the resulting charge distribution is obtained from
\begin{equation}
\label{eq:msr}
\langle r^2 \rangle = \frac{1}{Ze} \int_0^\infty \!\!\!\! dr\ r^2 \rho_{ch}(r) 
\end{equation}
and has been employed to constrain the nonlocal HF potential to generate good agreement with the experimental mean square radius of the ${}^{40}$Ca charge distribution.
The parameters in Table~\ref{Tbl:parms} generate a value of 3.45 fm compared to the experimental result of 3.45 fm taken from the Fourier-Bessel analysis given in Ref.~\cite{deVries1987}.
\begin{figure}[tbp]
\includegraphics*[ width=.48\textwidth,angle=-90]{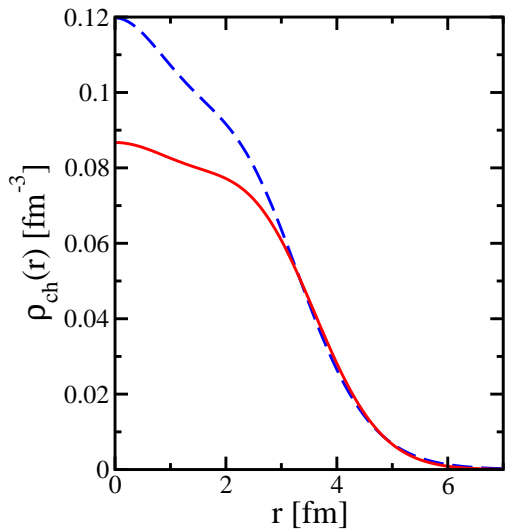}
\caption{(Color online) Experimental charge density of ${}^{40}$Ca~\protect\cite{deVries1987} (solid) compared with the DOM result (dashed).}
\label{fig:chd}
\end{figure}
We  compare the calculated charge density with the experimental one in Fig.~\ref{fig:chd}.
It is obvious that there is still a significant discrepancy with the experimental charge density near the origin which requires further analysis.
Before addressing this issue in more detail we remark on the individual radii of the quasihole orbits and compare these with results obtained from the (\textit{e,e}$^{\prime }$\textit{p}) reaction.
By constraining the nonlocal HF potential to reproduce the mean square charge radius, we obtain radii of individual orbits that are somewhat smaller than for the local DOM fit.
We list the radii for the solutions of Eq.~(\ref{eq:DSeq}) (without the imaginary part) in Table~\ref{Tbl:radii} and compare the DOM results with each other.
\begin{table}[tbp]
\caption{Radii for proton orbits in ${}^{40}$Ca for the local and nonlocal DOM implementation.}
\label{Tbl:radii}%
\begin{ruledtabular}
\begin{tabular}{ccc}
orbit  &local [fm] &nonlocal [fm] \\
\hline
$0s_{1/2}$ & 2.34 & 2.36 \\
$0p_{3/2}$ & 2.99 & 2.92 \\
$0p_{1/2}$& 2.98 & 2.90 \\
$0d_{5/2}$ & 3.54 & 3.36\\
$1s_{1/2}$ & 3.87 & 3.60 \\
$0d_{3/2}$ & 3.71 & 3.52 
\end{tabular}
\end{ruledtabular}
\end{table}
We first note that the radii listed are for point particles and become larger when folded with the proton charge density to generate the result for the nonlocal case that agrees with the experimental mean square radius of the charge distribution.
This constraint makes the nonlocal radii smaller than the ones for the local DOM.
The radii for the local DOM are in good agreement with the (\textit{e,e}$^{\prime }$\textit{p}) analysis of Refs.~\cite{Kramer89,Kramer01} for the valence hole states.
A direct comparison of the point proton radii is appropriate, since the analysis of the data employs bound-state wave functions for point nucleons but includes the coupling to an extended charge distribution of the proton by employing the off-shell $(e,p)$ cross section.
We do note that the (\textit{e,e}$^{\prime }$\textit{p}) analysis generates a sp wave function with respect to the $A-1$ system.
This is not the case for the calculation of the DOM propagator with the nonlocal HF potential.
Since the DOM calculation generates proton scattering wave functions in a wide energy domain and also calculates the proton overlap function for removal to valence hole states, it provides all the ingredients that are employed in the analysis of the (\textit{e,e}$^{\prime }$\textit{p})  cross sections.
Such cross sections, in turn, can therefore be used in future DOM calculations to constrain the nonlocal potential to check whether the radii obtained for the nonlocal DOM describe the data.
Indeed, we note that elastic-nucleon-scattering data only determine the phase shifts associated with the asymptotic scattering wave functions and do not provide strong constraints on the interior scattering waves.

The possibility of generating the one-body density matrix from a nonlocal DOM calculation provides access to the natural orbits of the system.
We discuss these orbits before we address the discrepancy of the DOM charge density with the experimental one, since the shape of the natural orbits provides some initial clues as to what is missing in the DOM potentials as implemented so far.
By diagonalizing the one-body density matrix given in Eq.~(\ref{eq:dmat}) one obtains the natural orbits for each $\ell j$ combination together with the corresponding occupation numbers.
It is therefore possible to write 
\begin{equation}
n_{\ell j}(r,r') = \sum_{n} n^{no}_{n \ell j} \psi^{no^*} _{n \ell j}(r) \psi^{no} _{n \ell j}(r')  ,
 \label{eq:norb}
 \end{equation}
 with $n^{no}_{n \ell j}, \psi^{no}_{n \ell j}(r)$ the corresponding occupation numbers and wave functions for the natural orbits.
We note that these wave functions are normalized to unity.
While there is a correspondence for $n$-values that are nominally occupied between the natural orbits with large occupation numbers and overlap functions that correspond to mostly occupied states, this is lost for the natural orbits with small occupation numbers~\cite{Polls95}.
Indeed, natural orbits with small occupation numbers are more confined than those with large occupation and have no relation with mostly empty orbits associated with, for example, DOM potentials. 

\begin{figure}[tbp]
\includegraphics*[width=.4\textwidth,angle=-90]{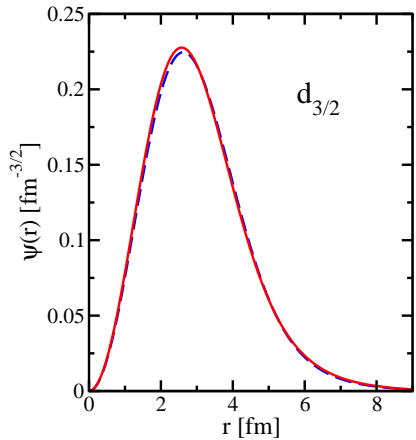}
\caption{(Color online) Comparison of the wave functions for the $d_{3/2}$ natural orbit (dashed) and the corresponding quasihole result (solid). The quasihole wave function has a slightly larger radius.}
\label{fig:wd32}
\end{figure}

\begin{figure}[bp]
\includegraphics*[width=.4\textwidth,angle=-90]{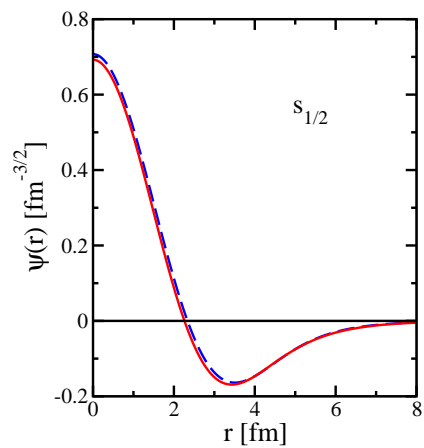}
\caption{(Color online) As in Fig.~\ref{fig:wd32} but for proton $s_{1/2}$ orbit with one node.}
\label{fig:ws12a}
\end{figure}

\begin{figure}[tbp]
\includegraphics*[width=.4\textwidth,angle=-90]{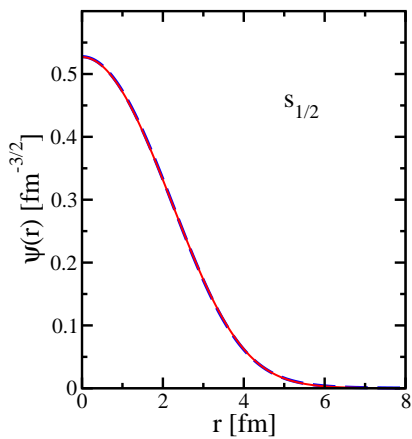}
\caption{(Color online) As in Fig.~\ref{fig:wd32} but the proton $s_{1/2}$ orbit with no node.}
\label{fig:ws12b}
\end{figure}

We compare in Fig.~\ref{fig:wd32} the proton $d_{3/2}$ natural orbit (solid) with the corresponding quasihole wave function (dashed) obtained by diagonalizing Eq.~(\ref{eq:DSeq}) at the correct energy.
Both wave functions are normalized to 1 and are basically indistinguishable.
This feature was also observed in the microscopic calculation of Ref.~\cite{Polls95} for ${}^{16}$O.
For the $s_{1/2}$ orbit two natural orbits are generated with large occupation numbers.
In this case the comparison in Figs.~\ref{fig:ws12a} and \ref{fig:ws12b} shows quite similar behavior.
The DOM quasihole wave functions exhibit the expected shell-model wave function characteristics since they have been generated by solving Eq.~(\ref{eq:DSeq}) and occur at quite different energies.
The natural orbits are generated quite differently as they require an integration of the spectral density over all energies up to the Fermi energy with a subsequent diagonalization of the the one-body density matrix without any direct reference to a Schr{\"{o}}dinger-like equation.
Nevertheless, the wave functions of natural orbits with large occupation numbers appear almost indistinguishable from their quasiparticle counterparts.

Occupation numbers for natural orbits are collected in Table~\ref{Tbl:noocc}.
\begin{table*}[btp]
\caption{Occupation numbers of natural orbits.}
\label{Tbl:noocc}%
\begin{ruledtabular}
\begin{tabular}{cccccccc}
n  &$s_{1/2}$ & $p_{3/2}$ & $p_{1/2}$ & $d_{5/2}$ & $d_{3/2}$ & $f_{7/2}$ & $f_{5/2}$ \\
\hline
\\
1 & 0.926 & 0.921 & 0.905 & 0.899 & 0.858 & 0.109 & 0.064\\
2 & 0.881 & 0.072 & 0.062 & 0.037 & 0.032 & 0.024 & 0.020\\
3 & 0.032 & 0.021 & 0.020 & 0.015 & 0.014 & 0.010 & 0.010\\
4 & 0.015 & 0.010 & 0.009 & 0.007 & 0.007 & 0.006 & 0.005\\
5 & 0.007 & 0.005 & 0.005 & 0.004 & 0.004 & 0.003 & 0.003\\
$\sum_n$ & 1.86 & 1.03 & 1.00 & 0.96 & 0.92 & 0.15 & 0.10 \\
\end{tabular}
\end{ruledtabular}
\end{table*}
The number of large eigenvalues (comparable to unity) corresponds exactly to the expected number of the simple shell model.
All other eigenvalues are small and are associated with wave functions with increasing number of nodes.
We also include the sum of the occupation numbers for each $\ell j$ combination in Table~\ref{Tbl:noocc}.
Somewhat surprising is that the largest deviation occurs for the $s_{1/2}$ orbit (in that case from 2, since there are nominally two levels occupied).
As observed in Ref.~\cite{Polls95}, the largest eigenvalues for the nuclear natural orbits are substantially larger than the corresponding one for drops of a finite number of ${}^{3}$He atoms~\cite{Lewart88}.
Since short-range correlations associated with the underlying bare interaction are included in the work of Refs.~\cite{Polls95} and \cite{Lewart88}, this difference is mostly related to the much stronger repulsion between ${}^{3}$He atoms which \textit{e.g.} in the liquid at saturation leads to a depletion of the Fermi sea of more than 50\%~\cite{Mazz04}. 
Nucleon-nucleon interactions typically generate 10-15\% depletion due to short-range and tensor correlations~\cite{Rios09}.
The inclusion of short-range correlations is partly accomplished by the assumed energy dependence of the volume term of the imaginary part of the DOM potential which is based on nuclear matter calculations~\cite{Muther05}.
Such an imaginary term is responsible for the global depletion of orbits including those that are deeply bound.
We note that the complementary admixture of high-momentum components is not yet incorporated by the current DOM implementation, as will become clear in the following discussion.
We also observe that the large occupation numbers of the natural orbits calculated for ${}^{16}$O in Ref.~\cite{Polls95} are 5-10\% larger than the DOM numbers quoted in Table~\ref{Tbl:noocc}.
It appears reasonable to interpret this difference to be due to the proper inclusion of low-energy, and therefore long-range correlations that are represented by the surface components of the DOM potentials.
The imaginary part of the DOM potential up to about 50 MeV is dominated by the surface potential.
The associated absorption is well-constrained by the differential cross sections for elastic nucleon scattering in this energy domain.
It is then also unavoidable that sp strength from below the Fermi energy is removed to this energy domain when such potentials are employed in the solution of the Dyson equation.

DOM implementations usually generate occupation numbers for quasihole and other bound orbits~\cite{Mahaux91} based on approximate expressions. 
It is therefore useful to compare these with the more accurate results obtained by integrating the corresponding strength up to the Fermi energy.
This can also be accomplished by using the quasihole wave functions and performing an integration over $r$ and $r'$ involving the one-body density matrix.
For the proton $d_{3/2}$ quasihole orbit we find an occupation number of 0.86 which is identical
to the one for the natural orbit.
This is hardly surprising since these wave functions are almost identical as shown in Fig.~\ref{fig:wd32}.
The DOM result of Ref.~\cite{Charity07} yields 0.82 reasonably close to the Green's function result of the nonlocal DOM.
For the $1s_{1/2}$ quasihole state calculated at -9.8 MeV we obtain an occupation number of 0.88 a little larger than the result obtained with the local DOM which generates 0.85.
For the $0s_{1/2}$ state the corresponding numbers are 0.93 (nonlocal) and 0.93 (local) in complete agreement.
This suggests that the corresponding approximate expression in the local DOM may be appropriate for the occupation numbers deep in the Fermi sea.
It is however well known that for such orbits unreasonable spectroscopic factors can be generated.
For the $0s_{1/2}$ orbit the local DOM yields for example an unphysical spectroscopic factor of 1.11 (see Table~\ref{Tbl:spfac}).
The corresponding calculation for the nonlocal DOM yields 0.98, a not unphysical result but also not useful since it is larger than the properly calculated occupation number of 0.93.
This is hardly surprising since the calculation employing Eq.~(\ref{eq:sfac}) neglects the role of the imaginary part of the self-energy completely.
As discussed earlier, it is therefore only useful to consider spectroscopic factors near the Fermi energy where the imaginary part of the self-energy is insignificant or zero.
At other energies, it is more appropriate to consider the spectral function as illustrated in Fig.~\ref{fig:allsf}.

\begin{figure}[tbp]
\includegraphics*[width=.4\textwidth,angle=-90]{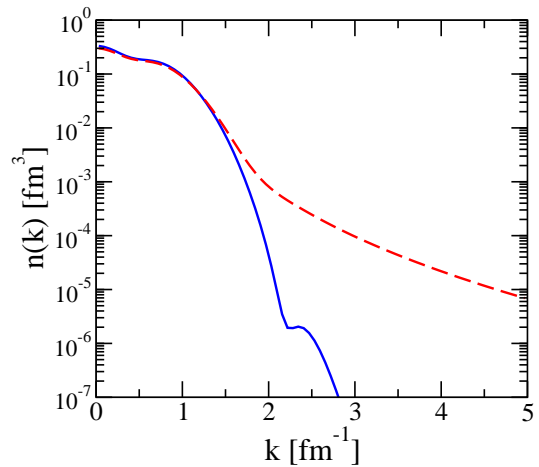}
\caption{(Color online) Comparison of the total momentum distribution calculated according to Eq.~(\ref{eq:tmomd}) (dashed) with the one obtained from the quasihole contributions (solid).}
\label{fig:tmom}
\end{figure}
\begin{figure}[tbp]
\includegraphics*[width=.4\textwidth,angle=-90]{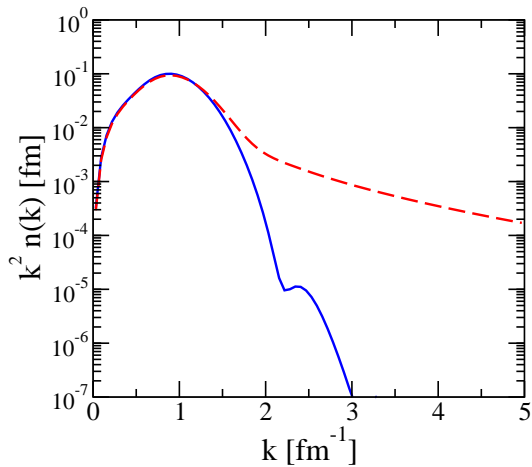}
\caption{(Color online) Momentum distribution for protons as in Fig.~\ref{fig:tmom} but weighted by $k^2$.}
\label{fig:wtmomd}
\end{figure}

As the discussion of natural orbits has shown, the effect of short-range and tensor correlations are only included in so far as orbits below the Fermi energy are depleted by the presence of a substantial imaginary part at large positive energy associated with the volume contribution to the DOM self-energy.
The associated presence of high-momentum components is not explicitly guaranteed and we will analyze this in the following.
The total proton momentum distribution for protons (normalized by $Z$) is obtained from
\begin{equation}
n(k) = \frac{1}{Z}\sum_{\ell j} (2j+1) n_{\ell j}(k) .
\label{eq:tmomd}
\end{equation}
We obtain the partial momentum distributions $n_{\ell j}(k)$ by first generating the momentum-space spectral function by performing a double Fourier-Bessel transform of the spectral density in coordinate space
\begin{equation}
S_{\ell j}(k;E) = \frac{2}{\pi^2} \!\! \int_0^\infty \!\!\!\!\!\! dr r^2 \!\!\! \int_0^\infty \!\!\!\!\!\! dr' r'^2 j_{\ell}(kr) \mbox{Im} G_{\ell j}(r,r';E) j_\ell(kr') .
\label{eq:kspecf}
\end{equation} 
The momentum distribution for a given $\ell j$ is then obtained from
\begin{equation}
n_{\ell j}(k) = \int_{-\infty}^{\varepsilon_F} dE\ S_{\ell j}(k;E) .
\label{eq:momd}
\end{equation}
In Fig.~\ref{fig:tmom} we display the total proton momentum distribution by the dashed line.
For comparison we also show the momentum distribution from the quasihole wave functions (normalized to one) by the solid line.
As discussed in previous work (see \textit{e.g.} Refs.~\cite{Muther94,Muther95,Polls95}),
these quasihole contributions are mostly associated with wave functions near the Fermi energy and hardly contain any high-momentum components.
The presence of high-momentum components is demonstrated by the dashed line in Fig.~\ref{fig:tmom}.
We emphasize their contribution by showing in Fig.~\ref{fig:wtmomd} the momentum distribution weighted by $k^2$. 
Interestingly we find about 10\% of the protons actually have momenta beyond 1.4 fm$^{-1}$.
This number is in reasonable agreement with the 10\% generated for ${}^{16}$O in the calculations of Refs.~\cite{Muther94,Muther95,Polls95}.
These calculations generate high-momentum components that are in quite good agreement in the aggregate with the results of Ref.~\cite{Rohe04}.
It is therefore clear that the present version of the DOM includes sufficient flexibility to represent these experimentally well established ingredients at least in aggregate.

\begin{figure}[bp]
\includegraphics*[width=.4\textwidth,angle=-90]{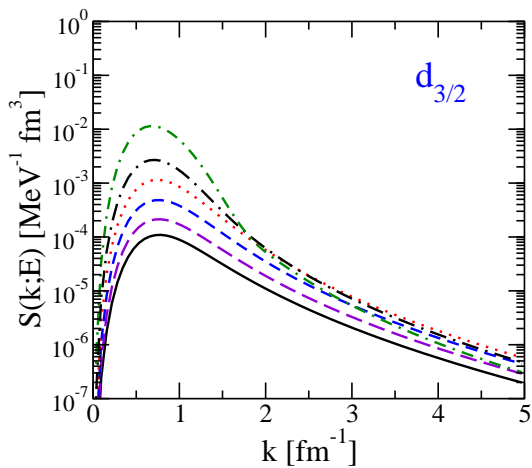}
\caption{(Color online) Momentum-space spectral function for $d_{3/2}$ quantum numbers at different energies. The highest curve is obtained at -25 MeV and each successive lower curve (at small momenta) represents a 25 MeV step lower in energy with the last curve representing the spectral function at -150 MeV.}
\label{fig:kd32}
\end{figure}

\begin{figure}[tbp]
\includegraphics*[width=.4\textwidth,angle=-90]{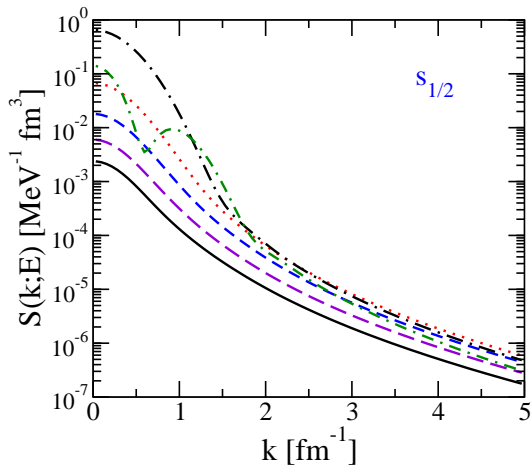}
\caption{(Color online) Momentum-space spectral function for $s_{1/2}$ quantum numbers at different energies as explained in the text.}
\label{fig:ks12}
\end{figure}

Looking in more detail we note that the expected behavior of high-momentum components, \textit{i.e.} increasing importance with increasing separation energy, is not contained in the DOM spectral functions.
We illustrate this observation in Fig.~\ref{fig:kd32} by plotting the $d_{3/2}$ spectral function in momentum space at different energies; starting at -25 MeV with steps of 25 MeV all the way to -150 MeV.
The figure illustrates that the \textit{shape} of the momentum content of the spectral function hardly changes as a function of energy, especially when momenta above 1.4 fm$^{-1}$ are considered.
This latter feature is completely opposite to the effect expected of short-range correlations. 
As discussed at length in Refs.~\cite{Muther94,Muther95,Polls95}, the presence of high-momentum components becomes more pronounced with decreasing energy (away from the Fermi energy) unlike the results shown in Fig.~\ref{fig:kd32}.
The former result can be easily understood on the basis of simple considerations involving momentum conservation and the location of the relevant 2h1p states that are required for the admixture of high-momenta~\cite{Dickhoff04}.
We note that this effect has been experimentally confirmed by the absence of appreciable high-momentum components in the valence hole states in ${}^{208}$Pb~\cite{Bobel94}.
The results of Ref.~\cite{Rohe04} further illustrate that high-momentum components emerge with decreasing energy and dominate at energies substantially below the bottom of the traditional potential well used to describe mean-field nucleons.
In Fig.~\ref{fig:ks12} we display the momentum content for the $s_{1/2}$ orbit at the same energy values.
In this case the short-dash-dot curve at -25 MeV is reminiscent of the $1s_{1/2}$ quasihole wave function and the long-dash-dot curve at -50 MeV is close to the quasihole peak of the $0s_{1/2}$ orbit.
At lower energy this shape persists but the high-momentum content (apart from slowly decreasing)
exhibits no essential change in energy as for the $d_{3/2}$ channel.

In order to describe the correct behavior of the high-momentum components in the DOM it will be necessary in the future to make the geometry of the potential dependent on energy.
Indeed, by reducing the radius of the confining nuclear potential with decreasing energy, one may expect to raise the high-momentum content and generate the behavior predicted in Refs.~\cite{Muther94,Muther95} and experimentally confirmed in~\cite{Rohe04}.
Since the geometry of the DOM potential has been assumed independent of energy in the current implementations, this will increase the computational effort substantially since the application of the subtracted dispersion relation will have to be performed also as a function of the coordinates for which the real part of the dispersive part is required.
The work of Refs.~\cite{Muther94,Muther95} was performed in momentum space and it may be necessary to consider DOM implementations which rely on momentum-space formulations, at least as far as short-range correlations are concerned.

Having established some missing ingredients in the description of high-momentum components, we now argue that this has consequences for the description of the nuclear charge density.
As discussed in Ref.~\cite{MuSi04}, the role of short-range correlations is to remove some nuclear charge, present in the mean-field description in terms of the occupied $s_{1/2}$ states, from the origin to larger radii but not to the surface, which is dominated by quasihole contributions.
While some of this charge returns to the origin as partially occupied higher $s_{1/2}$ states, most of this strength is associated with higher $\ell$-values, similar to the results obtained in Refs.~\cite{Muther95,Polls95}.
It is therefore reasonable to expect that a proper treatment of short-range correlations with the attendant presence of high-momentum (higher $\ell$) components (constrained by the experimental data~\cite{Rohe04}) will make it possible to obtain an accurate fit to the nuclear charge density in a DOM framework.

It is well known that the sp propagator allows for the calculation of the energy per particle from the contribution of the underlying two-body interaction.
For the present case, it is useful to employ this result in momentum space.
The energy per proton of the ground state can, for example, be obtained by calculating~\cite{Dickhoff08}
\begin{eqnarray}
\frac{E({}^{40}\textrm{Ca})}{Z} =  \frac{1}{2Z} \sum_{\ell j} (2j+1) \int_0^\infty \!\!\! dk k^2 \frac{k^2}{2m} n_{\ell j}(k) \nonumber \\
+   \frac{1}{2Z} \sum_{\ell j} (2j+1) \int_0^\infty \!\!\! dk k^2 \int_{-\infty}^{\varepsilon_F} dE\ E S_{\ell j}(k;E) .
\label{eq;eperz}
\end{eqnarray}
For the present DOM potential we obtain only -2.91 MeV per proton which includes the effect of the Coulomb interaction.
A similar calculation for the neutrons yields -6.51 MeV per neutron for a total of -4.71 MeV per particle.
This result represents about 60\% of the experimental result.
This is a remarkable result since the spectral information and the location of the bound levels in combination with a considerable wealth of elastic scattering data is described by the DOM self-energy.
However, also in this case we can point to the lack of the correct description of high-momentum components that can resolve this issue.
In Ref.~\cite{Muther95} it was shown that the quasihole contribution to the energy per particle is about 35\% in ${}^{16}$O whereas 65\% is generated by the continuum contribution at large negative energies where high-momenta dominate.
This result is noteworthy also since only 10\% of the nucleons are considered to have high momenta as confirmed by experiment.
A similar situation appears to apply in the case of the DOM analysis of ${}^{40}$Ca.
Since the total number of high-momentum components appears reasonable, it appears that their appearance at more negative energy will be able to resolve part of the discrepancy for the total energy of the ground state.
It must also be noted that an important contribution from three-body forces may have to be considered.
It appears therefore reasonable to expect that all data that are not yet well reproduced at present, can be better described in a future DOM implementation which incorporates the contribution of about 10\% of high-momentum nucleons with the correct energy dependence.

\section{Summary and conclusions}

\label{Sec:conclusions} 
The present work aims at extending the DOM approach, which so far has been mostly applied to  describe elastic nucleon scattering, into the domain below the Fermi energy by employing additional experimental data to constrain the potentials.
By introducing an explicit nonlocal HF-like potential, it is possible to reinterpret the DOM potential as a nucleon self-energy when the corresponding nonlocality correction is implemented to generate the intended normalization of the DOM potential containing the local but energy-dependent HF contribution.
This procedure has been adopted for DOM potentials that were previously obtained for ${}^{40}$Ca~\cite{Charity07}. 
The possibility to interpret the DOM potential below the Fermi energy as the nucleon self-energy broadens the links with experimental data substantially.
The solution of the Dyson equation below the Fermi energy with this self-energy then leads to the nucleon sp propagator and the corresponding one-body density matrix.
A Perey-Buck type nonlocal potential was chosen to represent the HF potential.
Its parameters were chosen to describe the energies of the valence hole states and the mean square radius of the charge-density distribution.
The latter feature illustrates the new possibility to constrain the DOM potential by data that pertain to information associated with one-body properties of the nuclear ground state, since the sp propagator provides this access for any one-body operator.
Various quantities that are obtained from approximate expressions with the usual local form of the DOM potential are compared with the nonlocal solution.
Spectroscopic factors near the Fermi energy appear to be stable quantities, but are no longer useful for deeply bound states.
Instead, we advocate the construction of the complete spectral function in particular when comparison with nucleon knockout experiments are considered.
The nonlocal HF potential also limits the binding of the lowest $s_{1/2}$ orbital in agreement with corresponding experimental information. 
This orbit tends to become too deeply bound with the local version of the DOM potential.

The diagonalization of the one-body density matrix allows for the study of natural orbits and associated occupation numbers (eigenvalues). 
Results are qualitatively similar to microscopic calculations performed earlier for ${}^{16}$O and drops of a finite number of ${}^3$He atoms.
For each orbit that is filled in a simple mean-field picture, there is a corresponding large eigenvalue, while all other eigenvalues for this $\ell j$ combination are at least an order of magnitude smaller.
Independent of whether one or two such orbits are occupied below the Fermi energy, like for the $d_{3/2}$ and $s_{1/2}$, respectively, there is little difference between the quasihole (overlap) and natural orbit wave functions.
Although the mean square radius of the charge distribution agrees with the experimental value, a comparison with the complete density distribution shows that too much charge is calculated near the origin.
By studying the momentum content of the spectral function and associated momentum distribution, we observe that current DOM potentials generate about 10\% high-momentum components in agreement with experimental observations~\cite{Rohe04} for light nuclei.
Microscopic calculations of short-range correlations however demonstrate that high-momentum components become increasingly important with decreasing energy away from the Fermi energy, which is also confirmed by experiment.
This feature must be represented by a change in geometry of the potential with decreasing energy, which is currently not present in DOM potentials but can be implemented in future applications.
Since short-range correlations remove charge from the origin ($s_{1/2}$ orbits) and place it at larger radii although not at the surface~\cite{MuSi04}, it is expected that the inclusion of high-momentum components in the DOM potentials will allow a better description of the complete charge distribution.
Similar considerations suggest that these high-momentum components will also generate a substantially improved energy per particle.

Since the DOM generates both scattering wave functions at positive energy as well as quasihole overlap function, it will in the future also be possible to describe (\textit{e,e}$^{\prime }$\textit{p}) cross sections directly and use these data to constrain the DOM potential further.
At present, the indirect comparison with spectroscopic factors derived from these data suggests no inconsistency.
Nevertheless, it is important to explore the interior scattering wave function of the outgoing proton in this process, since it may be sensitive to the nonlocal features of the potential.

\acknowledgments This work was supported by the U.S. National Science Foundation under grants PHY-0652900 and PHY-0968941 and the U.S. Department of Energy,
Division of Nuclear Physics under grant DE-FG02-87ER-40316.

\bibliography{level_de}

\end{document}